\documentclass[aps,prd,preprintnumbers,groupedaddress,nofootinbib,amssymb,eqsecnum,notitlepage]{revtex4}
\usepackage{amsfonts}
\usepackage{graphicx}  
\usepackage{dcolumn}   
\usepackage{bm}        
\usepackage{amssymb}   
\usepackage{amsmath}
\usepackage{mathtools}
\usepackage{cases}
\usepackage{comment}
\usepackage{color}
\usepackage{epsfig}
\usepackage[normalem]{ulem}
\usepackage{amssymb,amsmath,bm,bbm}
\usepackage{verbatim}
\usepackage{appendix}
\newcommand{\be}{\begin{equation}}
\newcommand{\ee}{\end{equation}}
\newcommand{\bea}{\begin{eqnarray}}
\newcommand{\eea}{\end{eqnarray}}

\newcommand{\blue}[1]{{\color{blue} #1}} 
 
\newcommand{\my}[1]{{\color{red} #1}}

\begin{document}

\title{Reconstruction of Primordial Power Spectrum of curvature perturbation from the merger rate of Primordial Black Hole Binaries}

\author{Rampei Kimura$^{1}$,
Teruaki Suyama$^{2}$,
Masahide Yamaguchi$^{2}$ and 
Ying-li Zhang$^{3,4}$}

\affiliation{$^1$Waseda Institute for Advanced Study, Waseda University, 1-6-1 Nishi-Waseda, Shinjuku, Tokyo 169-8050, Japan\\
$^2$Department of Physics, Tokyo Institute of Technology,
2-12-1 Ookayama, Meguro-ku, Tokyo 152-8551, Japan\\
$^3$School of Physics Science and Engineering, Tongji University, Shanghai 200092, China\\
$^4$Institute for Advanced Study of Tongji University, Shanghai 200092, China}

\date{\today}

\begin{abstract}
The properties of primordial curvature perturbations on small scales are still unknown while those on large scales have been well probed by the observations of the cosmic microwave background anisotropies and the large scale structure. In this paper, we propose the reconstruction method of primordial curvature perturbations on small scales through the merger rate of binary primordial black holes, which could form from large primordial curvature perturbation on small scales.
\end{abstract}


\maketitle

\section{Introduction}

The idea of Primordial Black Holes (PBHs) can be traced back to 1960s, which was firstly investigated by Zel'dovich and Novikov~\cite{Zeldovich:1967}. Later, in 1971, Hawking pointed out that in the primordial universe, black holes (BHs) can be formed by the gravitational collapse of a highly overdense region of inhomogeneities~\cite{Hawking:1971ei}. Since then, the studies of PBHs have attracted a lot of attention mainly in two aspects. 
Firstly, since PBHs only interact via gravity, without introducing new physics beyond the standard model, they are natural candidates for dark matter (DM) which comprises 25$\%$ of the critical density (the earliest idea can be found in 1970~\cite{Chapline:1975ojl}).  Most of the constraints on the fraction of PBHs in dark matter, come from the gravitational lensing and dynamical constraints. However, recent reanalysis opened a parameter window where PBHs with mass $10^{-16}-10^{-10}M_{\odot}$ can constitute all of the DM~\cite{Niikura:2017zjd,Katz:2018zrn,Smyth:2019whb,Montero-Camacho:2019jte}. This refreshes the interest of the investigation of PBHs as the origin of DM~\cite{Pi:2017gih,Cai:2018tuh}. 
Secondly, the detected binary black hole (BBH) mergers during the past O1, O2 and first half of O3 runs of the Laser Interferometer Gravitational-wave Observatory (LIGO)~\cite{Abbott:2016blz,Abbott:2016nmj,TheLIGOScientific:2016pea,Abbott:2017vtc,Abbott:2017oio,Abbott:2017gyy,LIGOScientific:2018mvr,Venumadhav:2019lyq,Abbott:2020niy}  have shown two unexpected characteristics of binary black hole (BBH): the high masses (larger than 20 $M_\odot$) and low effective spin. The origin of these heavy BHs and the formation of such BBHs which merge within the age of the Universe are still in debate~\cite{Belczynski:2010tb,Dominik:2012kk,Dominik:2013tma,Dominik:2014yma,Belczynski:2016obo} (for a comprehensive summarization, see e.g.~\cite{TheLIGOScientific:2016htt,Miller:2016krr}). One of the possible explanations is that the mergers could be primordial in origin~\cite{Bird:2016dcv,Clesse:2016vqa,Sasaki:2016jop,Sasaki:2018dmp, Garcia-Bellido:2020pwq}.

Up to now, most of works concentrate on predicting the PBH construction and its associated BBH mergers for (given) primordial curvature perturbations. However, in near future, we expect to detect much more BBH mergers. In this new era, the following (opposite) question will become more and more important: How can we reconstruct the features of primordial curvature perturbations such as power spectra and bispectra from the observations of BBH mergers? In fact, if (some of) these BBH mergers could be attributed to PBHs, we will be able to reconstruct the features of primordial curvature perturbations which would predict PBHs and then such PBHs are responsible for the detected BBH mergers. This reconstruction program is quite important because, over the past two decades, on cosmological scales (larger than Mpc scale), the observations of the cosmic microwave background (CMB) anisotropies and the large scale structure indicate an almost scale-invariant primordial curvature power spectrum with accurately measured amplitude and tilt~\cite{Spergel:2003cb,Aghanim:2018eyx,Akrami:2018odb}. However, on small scales (smaller than Mpc scale), we still lack the information of primordial curvature power spectrum though some upper bounds are obtained~\cite{Chluba:2019kpb,Jeong:2014gna,Nakama:2014vla,Inomata:2016uip,Allahverdi:2020bys}. This reconstruction program could directly determine the information of primordial curvature perturbations on small scales. As far as we know, no such attempt has yet been done albeit its importance.

In this paper, we will organize such a reconstruction program. 
Our finding is that the reconstruction program must go through three distinct steps to successfully reconstruct the primordial power spectrum.
In step (i), one needs to relate the feature of primordial curvature perturbations (such as power spectrum) to the (smoothed) density contrast (characterized by the mass variance for example). In step (ii), the PBH mass function needs to be related to the (smoothed) density contrast. In step (iii), the PBH mass function needs to be related to the (would-be) observed rates of BBH mergers associated with PBHs. In each step, there should be one-to-one correspondence for this reconstruction program to work. 
As we will explain in details later,
each step suffers from its own theoretical uncertainties.
They stem from i)very rare and highly non-linear nature of the PBH formation which defeats a first-principle calculation and forces us to employ some phenomenological modeling where some ambiguities come into play,
and ii)our ignorance of the model of the early universe.
As a result, it is inevitable to adopt some assumptions to achieve the reconstruction of the primordial power spectrum.
In the next section, we will explain those assumptions one by one.
For the technical assumption, we will also discuss how we should change the reconstruction procedure when the assumption is relaxed.

This paper is organized as follows. In Section~\ref{assumptions}, we state the underlying four assumptions for the calculation and describe the motivations for these assumptions. In Section~\ref{principle}, we divide the reconstruction procedure into three steps. In Section~\ref{s1}, we relate the power spectrum of the Gaussian primordial curvature perturbations to the variance of the density perturbation smoothed by the top-hat window function. In Section \ref{s2}, we discuss the correspondence between the variance and the PBH mass function with the critical collapse phenomena being taken into account. In Section~\ref{s3}, by adopting the most updated formula of the merger rate, we finally show how to relate the PBH mass function to the (would-be) observed rates of BBH mergers associated with PBHs, assuming some of the observed binary BH merger events are attributed to PBH mergers, and show the PBH mass function can be reconstructed by putting these three steps together once the merger rate with sufficient mass resolution is determined. As an example, in Section~\ref{example}, we consider the simplest case where the effects of critical collapse and suppression factor are neglected and then find the one-to-one correspondence between the mass function and the merger rate in an analytic way. In Section~\ref{conclusion}, we draw the conclusion of this paper and give some discussions for further extension. In Appendix~\ref{appA}, for completeness, we make a general mathematical discussion on the one-to-one correspondence between the variance and the primordial power spectrum with an arbitrary window function.

\section{Underlying assumptions}\label{assumptions}

In this section, we state the following four assumptions that will be used in the next section:
\begin{description}
  \item[(0)] 
  In the future, some of the observed binary BH merger events
  are attributed to mergers of PBHs formed, and the merger rate distribution of the PBHs in the mass plane is observationally determined.
  \item[(1)]
  PBHs were formed out of rare high-$\sigma$ peaks of the primordial curvature perturbations in radiation dominated era.
  \item[(2)] The window function takes the {\it top-hat} form in $k$-space;
  \item[(3)] 
  The primordial curvature perturbation follows {\it Gaussian} distribution and the effect of non-linearity between the curvature
  and the density perturbations is ignored;
\end{description}

\underline{About the assumption (0)}\\
The origin of the BHs detected by LIGO/Virgo experiments remains an open question.
Broadly, there are two astrophysical channels of the BBH formation which could possibly explain the observed merger events
(for instance, see \cite{Mapelli:2018uds}).
One is the isolated binary formation channel in which the massive stars in a binary evolve into the BH binary
after experiencing the common envelope phase through which the binary shrinks its orbit and ends up with a tightly bound system. 
The other is the dynamical formation channel in which the two BHs form a binary by dynamical encounters
which occurs in dense environments such as globular clusters.
Another intriguing possibility is the PBH scenario, which could explain not only the observed binary black hole (BBH) merger events, 
but also the fact that many observed BBHs have large mass and low spin \cite{Abbott:2020niy}.
Given the large theoretical uncertainties in predicting the merger rate distribution of the individual astrophysical
channels, it is a challenging task to determine how much the PBHs contribute to the observed merger rate distribution 
from the data (see \cite{Kocsis:2017yty, Liu:2018ess, Chen:2019irf, Garriga:2019vqu, Clesse:2020ghq, Hall:2020daa, Wong:2020yig} as recent works along this direction).
In order to achieve this, we would ultimately need to reduce the theoretical uncertainties of the astrophysical channels and
subtract the astrophysical contributions from the observationally determined merger rate distribution
to identify the PBH signal.
Whether the merger rate distribution obtained in such a way is consistent with the mergers of PBHs 
can be checked by using the consistency relation, which is briefly discussed in the last section in this paper (see \cite{Kocsis:2017yty,Liu:2018ess,Garriga:2019vqu} as relevant previous studies).
Alternatively, the issue of discriminating from the astrophysical BHs can be circumvented if merger events at
cosmological redshifts $z\gtrsim 20$ \cite{Nakamura:2016hna} or of sub-solar mass BHs are detected since astrophysical BHs are thought 
to be very rare at such high redshifts and heavier than the solar mass (see \cite{Authors:2019qbw} which conducted search for sub-solar PBH mergers.).
In these cases, it would be straightforward to observationally determine the PBH merger rate distribution.
In this paper, we simply adopt the assumption (0) and suppose that one could know the merger rate of BBH originating from the PBHs because the main purpose of this paper is to give the method of the reconstruction of primordial curvature perturbations from such merger rates.
\\

\underline{About the assumption (1)}\\
Though various mechanism to produce PBHs such as cosmic topological defects~\cite{Hawking:1982ga,Crawford:1982yz,Hawking:1987bn,Moss:1994iq}, interacting dark matter clumps~\cite{Shandera:2018xkn} have been proposed so far, one of the most interesting possibilities is large curvature perturbations generated during inflationary expansion~\cite{Ivanov:1994pa,GarciaBellido:1996qt,Bullock:1996at,Ivanov:1997ia,Cai:2019jah,Palma:2020ejf,Cheong:2019vzl,Zhou:2020kkf,Chen:2020uhe}. In fact, through this mechanism of large curvature perturbations, a lot of works have been done to ``predict" PBHs, which might account for dark matter, MACHOs, (unidentified) cosmic rays, and so on. In these works, given a specific model of inflation, primordial curvature perturbations were predicted, and then, the formation of PBHs originating from their designed power spectra has been discussed. 
In principle, it is possible that early matter dominated era existed prior to the radiation dominated and PBHs were formed in such matter dominated era. Formation process of the PBHs in such era is very different from that in the radiation dominated era \cite{Khlopov:1980mg, Harada:2016mhb}.  
Given that the PBH formation in the radiation dominated era is the most standard possibility considered in the literature, we assume that PBHs originate from the primordial curvature perturbations in the radiation dominated era.
\\

\underline{About the assumption (2)}\\
The assumption (2) is imposed in order to obtain the analytic relation between the primordial power spectrum and the variance (see~\eqref{relation-s-p}). Of course there is no physical principle to choose a certain form of the window function. It is evident that different form of window function will change the final result, but it has been shown in~\cite{Ando:2018qdb,Young:2019osy} that the effect of the choice of window function will cause the uncertainty in the amplitude of the power spectrum only up to $\mathcal{O}(10\%)$. Hence, we concentrate on the top-hat form as a typical example in the main text. A numerical strategy to reconstruct the primordial power spectrum for other types of the window function is also briefly discussed in the appendix.\\

\underline{About the assumption (3)}\\
The primordial curvature perturbation is non-Gaussian in general. In the standard single-field slow-roll inflation models, the non-Gaussianity is suppressed by the slow-roll parameters. For other classes of inflation models, 
the level of non-Gaussianity as well as the types of non-Gaussianity are model-dependent. 
If the primordial curvature perturbations are non-Gaussian,
the PBH mass function and the PBH abundance depend not only on the power spectrum but also on higher-order statistics.
In other words, the reconstruction of the power spectrum from the observations of the PBH mergers does not work unless we make specific assumptions on the higher-order statistics.
The higher-order statistics will become important if the probability to realize the overdensity to form a PBH, which corresponds the tail of the probability density in realistic situation, differs significantly from the one given by the Gaussisan distribution.
For instance, for the curvature perturbation with the local-type non-Gaussianity parametrized by the $f_{\rm NL}$ parameter, 
the non-Gaussianity will change the PBH abundance from the Gaussian case if $|f_{\rm NL}| \gtrsim \sigma^2/\delta_{\rm th}^3$, where
$\sigma^2$ is the variance of the perturbation and $\delta_{\rm th}$
is the threshold for the PBH formation.
Since there is no standard non-Gaussian shape which is well-motivated than the other non-Gaussian shapes, we assume that the curvature perturbation is Gaussian.
Moreover, even if we assume the Gaussianity of the curvature perturbation, the non-linearity intrinsic to GR makes the density perturbation, which is relevant to the PBH formation, non-Gaussian.
We also neglect the effect of this non-Gaussianity for simplicity, 
which is the assumption (3).~\footnote{The effect of non-Gaussianity can be found in literatures, for example, Ref~\cite{Bullock:1996at,Ivanov:1997ia,PinaAvelino:2005rm,Lyth:2012yp,Byrnes:2012yx,Shandera:2012ke,Young:2013oia,Franciolini:2018vbk,Cai:2018dig,Yoo:2019pma,Atal:2019erb,Ezquiaga:2019ftu,Young:2014oea,Young:2015cyn,DeLuca:2019qsy}}. 
\\

The assumption (0) is mandatory (almost by definition) for the reconstruction program to be achieved successfully.
In this sense, this assumption is crucial in this paper.
The assumption (2) is not essential to the reconstruction although this is practically convenient from the technical point of view.
It is not clear if the violation of the assumption (1) and(or) the assumption (3) can spoil the reconstruction program.
Answer may depend on the early universe model for the PBH formation and the type of non-Gaussianity of the primordial curvature perturbations that we employ.
In that case, one needs to check the presence of the one-to-one correspondence in each step given below.
Addressing this issue is beyond the scope of this paper.
\\

\section{The principle of a reconstruction program of primordial curvature perturbations}\label{principle}
Under the assumptions (0)--(3), we describe the principle of the reconstruction program.
The logic chain of the reconstruction program is given as follows: 
\begin{align}\label{chain}
\mathcal{P}_{\mathcal{R}}(k)\stackrel{\rm(i)}{\Longleftrightarrow}\sigma^2(R)\stackrel{\rm(ii)}{\Longleftrightarrow} f(m)\stackrel{\rm(iii)}{\Longleftrightarrow}\mathcal{R}(m_1, m_2, t)\,,
\end{align}
where $\mathcal{P}_{\mathcal{R}}(k)$ is the power spectrum of the primordial curvature perturbations, $\sigma^2(R)$ is the variance of the density perturbation smoothed over a comoving length scale $R$, $f(m)$ is the PBH mass function, and $\mathcal{R}(m_1, m_2, t)$ is the merger rate density of the PBH binaries with the masses $m_1$ and $m_2$ at the cosmic time $t$. Their detailed definitions will be given later. 
In what follows, we explain these three steps one by one. \\

\subsection{Step~(i)}\label{s1}
     In radiation dominated universe, an initially overdense region collapses to a BH right after the horizon reentry if
     the average amplitude of the density contrast of the overdense region evaluated at the time of the horizon reentry 
     is larger than the threshold.
     In the literature, the density contrast $\Delta$ on the comoving slice is sometimes used as a useful quantity to 
     measure the PBH formation \cite{Young:2014ana}. 
     The density contrast smoothed over the comoving scale $R$ is defined by
     \begin{align}
     \Delta_R (t,{\bf x})=\int W({\bf y}-{\bf x};R) \Delta (t,{\bf y}) d{\bf y}.
     \end{align} 
     Here $W({\bf y};R)$ is the window function. Notice that $\Delta$ evolves as $\propto a^2$ on super-Hubble scales.
     What is relevant to the formation of PBH with mass $M$ is the variance of $\Delta_R$, 
     \begin{align}\label{sig}
     \sigma^2 (R)=\langle \Delta_R^2 (t_R,{\bf x}) \rangle =\int_0^\infty W^2(kR) \mathcal{P}_\Delta(t_R,k)~\mathrm{d}(\ln k),
     \end{align}
     where $R$ is determined as a function of $M$ by requiring
     that $M$ is the horizon mass evaluated at the time $t=t_R$ (i.e. $a(t_R) R=1/H(t_R)$) when the scale $R$ reenters the Hubble horizon, 
     and $W(kR)$ is the Fourier transformation of the window function.
     In order to connect $\sigma^2 (R)$ to ${\cal P}_{\cal R}(k)$, we use the relation between the power spectrum for the density contrast $\mathcal{P}_\Delta$ 
     and the one for the curvature perturbation during the radiation-dominated era given by
	$\mathcal{P}_{\mathcal{R}}$ as (see e.g., \cite{Green:2004wb})
	\begin{align}\label{Prelate}
	\mathcal{P}_\Delta(t, k)=\frac{16}{81}\left(\frac{k}{aH}\right)^4\mathcal{P}_{\mathcal{R}}(k).
	\end{align}	
	The right-hand side of this relation is the leading order in the Taylor expansion in the power of $k/(aH)$ 
	and higher order terms are ignored.
	We use this relation in evaluating the right-hand side of Eq.~(\ref{sig}), which will be a good approximation 
	since the unphysical contributions from the 
	sub-Hubble modes are suppressed thanks to the window function. Then, we have
	\begin{align}\label{sig2}
	\sigma^2 (R)=\frac{16}{81} \int_0^\infty W^2(kR) {(kR)}^4 {\cal P}_{\cal R}(k)~\mathrm{d}(\ln k).
	\end{align}

	Based on the assumption (2), we adopted
	the top-hat form window function in $k$-space
	\begin{align}\label{hat}
	W(kR)=\left\{\begin{matrix}
	\displaystyle \quad 1\quad\,; & 0<k<1/R\,,&~\\
	\\
	\displaystyle \quad 0\quad; & \mathrm{otherwise}\,,
	\end{matrix}
	\right.
	\end{align}
	
	Taking the derivative of \eqref{sig2} with respect to $1/R$, we then obtain
	\begin{align}\label{relation-s-p}
	\frac{81}{16} R^4\frac{{\rm d}}{{\rm d}R} \left( \frac{\sigma^2 (R)}{R^4} \right)  \bigg|_{1/R=k}=-k\mathcal{P}_{\cal R} (k).
	\end{align}
	 This is the desired relation which enables us to directly reconstruct the power spectrum of the curvature perturbation
	 $\mathcal{P}_{\mathcal{R}}(k)$ from (the derivative of) the variance $\sigma^2(R)$.
	 This reconstruction method works in an ideal case where $\sigma^2 (R)$ is determined by observations as a continuous 
	 function of $R$. 
	 In realistic observations, however, $\sigma^2 (R)$ will be measured only for some discrete values of $R$ unless we adopt
	 a specific fitting function for which free parameters entering the fitting function are fixed by observations.  
	 Furthermore, it is not clear how to extend the reconstruction method given above to other types of the window functions. 
	 Although the top-hat form in $k$-space is technically conveninent in the sense that it is straightforward to 
	 solve the integral equation (\ref{sig2}) for ${\cal P}_{\cal R}$,	
      this window function may not be considered as the most natural one from the physical point of view given that 
      the top-hat window function, when Fourier-transformed into the real space, is spatially extended beyond the Hubble horizon 
      and thus includes contribution of the perturbations outside the Hubble horizon which should not affect the PBH formation.
      In the appendix, we discuss about how to reconstruct ${\cal P}_{\cal R}$ when $\sigma^2(R)$ is given only discretely
      for a fixed window function which is not necessarily a top-hat form in $k$-space.

On the other hand, the variance $\sigma^2(R)$ can be uniquely determined from $\mathcal{P}_{\mathcal{R}}(k)$
by \eqref{sig2}.
Therefore, the step (i) in \eqref{chain} is reversible.
\\

\subsection{Step~(ii)}\label{s2}
Our next task is to reconstruct the variance $\sigma^2 (R)$ from the PBH mass function $f(M)$.
In this paper, the PBH mass function $f(M)$ is defined such that the quantity $f(M){\rm{d}}M$ 
represents the probability that a randomly chosen PBH is in the mass range $\left(M, M+{\rm{d}}M\right)$, 
%
\begin{align}\label{normf1}
  \int_0^\infty f(M) {\rm {d}}M = 1.
\end{align}

Computing the PBH mass function from the primordial perturbations has been a long-standing topic.
Since the formation of PBHs is a very rare and highly non-linear phenomena,  
there is no first principle derivation of the PBH mass function \footnote{One may think the PBH mass function
can be obtained by conducting the cosmological simulations which evolve the random primordial fluctuations initially 
defined on super-Hubble scales until all the interesting scales have reentered the Hubble horizon and by counting
the number of the produced PBHs and measuring the mass distribution.
However, since the PBH formation is very rare, the size of the simulation box must be extraordinarily large to contain
enough number of PBHs at the end of simulation. This is impossible at the current computational resource.}.
In the literature, semi-analytic formulations of the PBH mass function based on the numerical relativity computations of the PBH formation
in the spherically symmetric system and its phenomenological extrapolation to the more general 
random configuration without spherical symmetry have been addressed and refined repeatedly. 
In this paper, we adopt the mass function given in \cite{Byrnes:2018clq,Wang:2019kaf}
which connects the variance $\sigma^2 (R)$ with the PBH mass function $f(M)$
with the effect of the critical phenomena being included such that $M=KM_R\left(\Delta-\Delta_{\rm th}\right)^\gamma$. It is given by
\begin{align}
f(M)
&=\frac{1}{f_{\rm PBH}M} \int_{-\infty}^\infty 
\frac{K}{\sqrt{2\pi} \gamma \sigma (R)} \left(\frac{M_{\rm eq}}{M_R}\right)^{\frac12} {\left( \frac{M}{KM_R} \right)}^{1+\frac{1}{\gamma}}
\exp \left[ -\frac{1}{2\sigma^2 (R)} {\left( \Delta_{\rm th}+{\left( \frac{M}{KM_R} \right)}^{\frac{1}{\gamma}} \right)}^2 \right]
\frac{{\rm d} M_R}{M_R}\,,
\label{fm}
\end{align}
where $f_{\rm PBH}$ is the mass fraction of PBHs in dark matter, 
$M_{\rm eq}$ is the horizon mass at the matter-radiation equality epoch,
and the integration variable $M_R$ is the horizon mass at $t=t_R$,
\be\label{massR}
M_R=\frac{4\pi}{3} \rho (t_R) H^{-3}(t_R)=\frac{1}{2GH(t_R)},
\ee
by which $R$ is related to $M_R$.
This integration arises since the overdense region with fixed $R$ can collapse to the PBH of any mass 
in the range $0< M \lesssim M_R$
when the critical phenomena is taken into account.
$\gamma \approx 0.36$ is the critical exponent of the critical collapse, $\Delta_{\rm th}$ is the threshold
of the PBH formation, 
which depends both on the perturbation profile and the adopted window function \cite{Young:2019osy},
and $K=3.3$ is a numerical constant.

Eq.~(\ref{fm}) is a non-linear integral equation for $\sigma^2 (R)$ and it is impossible to find analytic solution. 
Furthermore, unfortunately, we were not able to find the numerical method to systematically reconstruct $\sigma^2 (R)$ for any given $f(M)$.
This is a purely mathematical problem and providing a full solution to this problem is beyond the scope of this paper.
Instead, here we sketch one ad hoc approach which may work sufficiently in some cases.
To this end, we first notice from Eq.~(\ref{fm}) that, from the physical point of view, 
the mass function $f(M)$ is expected not to decay faster than $M^{1/\gamma}$ as $M$ is decreased.
In order to see this, let us assume an extreme case where $\sigma^2 (R)=0$ for $R<R_*$.
In this case, the range of integration in Eq.~(\ref{fm}) is effectively restricted to $M_R>M_{R_*}$.
Thus, for $M \ll M_{R_*}$, Eq.~(\ref{fm}) becomes
\be
f(M)\approx\frac{M^{\frac{1}{\gamma}}}{f_{\rm PBH}} \int_{\ln M_{R_*}}^\infty 
\frac{K}{\sqrt{2\pi} \gamma \sigma (R)} \left(\frac{M_{{\rm eq}}}{M_R}\right)^{\frac12} {\left( KM_R \right)}^{-1-\frac{1}{\gamma}}
\exp \left( -\frac{\Delta_{\rm th}^2}{2\sigma^2 (R)} \right)
{\rm d} \ln M_R. \label{fm2}
\ee
The integrand is independent of $M$ and thus $f(M) \propto M^{\frac{1}{\gamma}}$
(For the case of the monochromatic power spectrum, this feature was derived in \cite{Niemeyer:1997mt} for the Gaussian perturbation
and was shown to hold for the general non-Gaussian perturbation as well in \cite{Yokoyama:1998xd}).
This tail $f(M) \propto M^{\frac{1}{\gamma}}$ reflects the fact that small-mass PBHs ($M\ll M_{R_*}$) are inevitably produced by the critical phenomena.
In this sense, this gives the minimum amount of PBHs even in the absence of the primordial perturbations at the scale corresponding to $M (\ll M_{R_*})$.
In reality, $\sigma^2 (R)$ will not be exactly zero for $R<R_*$,
and such additional perturbations will produce extra PBHs to the ones mentioned above,
which would only make $f(M)$ decrease slower than $M^{\frac{1}{\gamma}}$.
Thus, $f(M)$ is not expected to decay faster than $M^{\frac{1}{\gamma}}$ as we decrease $M$.
This consideration tells us that if the observed mass function turns out to decay faster than this limit,
it strongly disfavors the scenario of the PBH origin, which itself provides one universal test of the PBH scenario.
Since our main purpose in this subsection is the reconstruction of $\sigma^2 (R)$ from $f(M)$ under the assumption that the observed BHs 
have been robustly identified as PBHs and their mass function $f(M)$ has been well measured, we hereafter suppose
that the observed $f(M)$ is consistent with this limit.

\begin{figure}[t]
 \begin{center}
   \includegraphics[clip,width=10.0cm]{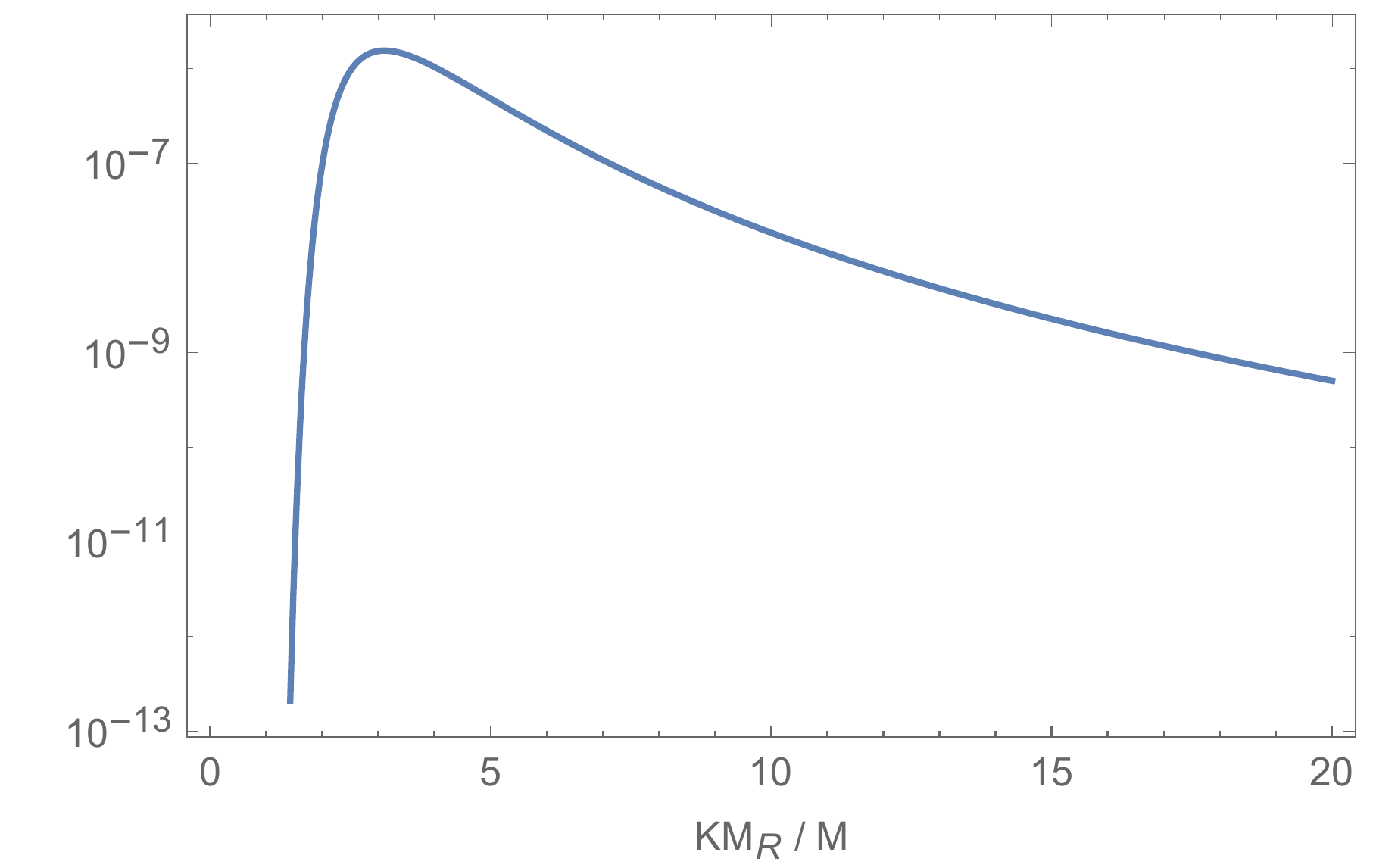}
   \caption{Integrand of Eq.~(\ref{fm}) as a function of $\frac{KM_R}{M}$ when $\sigma^2(R)$ is independent of $R$.
   We use $\gamma=0.36, \Delta_{\rm th}=0.4, \sigma^2=0.01$.
   Normalization of the vertical axis is chosen arbitrarily and only relative height is meaningful.}
   \label{fig-integrand}
 \end{center}
\end{figure}

If the observed $f(M)$ shows the scaling behavior $\propto M^{\frac{1}{\gamma}}$ in a certain mass range,
then the above argument suggests that $\sigma^2 (R)$ in the corresponding range of $R$ is negligibly small compared
to that for larger $R$. 
If, on the other hand, $f(M)$ is enhanced than the lower limit $M^{\frac{1}{\gamma}}$ for a wide mass range,
it implies that such PBHs were produced by nearly scale-independent $\sigma^2 (R)$ in the same range since $f(M)$ is
exponentially sensitive to $\sigma^2(R)$ (In any realistic situation, $\sigma^2 (R)$ must be much smaller than $\Delta_{\rm th}^2$
in order to avoid the overproduction of PBHs).
Fig.~\ref{fig-integrand} shows the integrand of Eq.~(\ref{fm}) as a function of $\frac{KM_R}{M}$.
We find that the integrand is sharply peaked at
\be
\frac{KM_R}{M} \approx \left( \frac{2\Delta_{\rm th}}{(2+5\gamma) \sigma^2 (R)} \right)^\gamma. \label{peak}
\ee
In the figure, the rapid fall-off on the left side is explained by the fact that the production of PBHs on super-Hubble scales does not occur,
and the decay on the right side is due to the critical phenomena.
Then, applying the method of the saddle point to Eq.~(\ref{fm}), we can approximately evaluate the integral as
\be\label{fmapp}
f(M) \approx f_{\rm PBH}^{-1} {\left( \frac{K}{M} \right)}^{\frac{3}{2}} \left(\frac{M_{\rm eq}}{\Delta_{\rm th}}\right)^{\frac{1}{2}} 
{\left[ \left(1+\frac52\gamma\right) \frac{\sigma^2 (R)}{\Delta_{\rm th}} \right]}^{\frac{1}{2}+\frac{3}{2}\gamma}
\exp \left( -\frac{\Delta_{\rm th}^2}{2\sigma^2 (R)} \right).
\ee
If $\sigma^2 (R)$ does not vary significantly, this gives an approximate relation between $f(M)$ and $\sigma^2 (R)$
(The scale $R$ for a fixed $M$ is determined by Eq.~(\ref{peak})).
At this level of approximation, it is clear that there is one-to-one correspondence between $f(M)$ and $\sigma^2 (R)$,
and $\sigma^2 (R)$ can be uniquely reconstructed from $f(M)$.

On the other hand, the non-critical collapse case corresponds to setting the constants $\gamma=0$ and $K\approx0.3$. In such case, starting from the first line of \eqref{fm}, the mass function can be expressed as
\begin{align}\label{fmnon}
f(M)\bigg|_{\rm non-critical}&=\frac{\sqrt{M_{\rm{eq}}}}{2f_{\rm PBH}}\left(\frac{K}{M}\right)^{\frac{3}{2}}~{\rm erfc}\left(\frac{\Delta_{\rm th}}{\sqrt{2}\sigma(R)}\right)\nonumber\\
&\approx\sqrt{\frac{M_{\rm eq}}{2\pi}}\frac{\sigma(R)}{f_{\rm PBH}\Delta_{\rm th}} {\left( \frac{K}{M} \right)}^{\frac{3}{2}}
\exp \left( -\frac{\Delta_{\rm th}^2}{2\sigma^2 (R)} \right)\,,
\end{align}
where ${\rm erfc}(x)$ is the complementary error function defined by
\begin{align}\label{errorori}
\mathrm{erfc}\left(x\right)\equiv\frac{2}{\sqrt{\pi}}\int_x^\infty e^{-z^2}{\rm{d}}z\,,
\end{align} 
and we used the asymptotic behavior of error function ${\rm erfc}(x)\approx(\sqrt{\pi}x)^{-1}\exp(-x^2)$ for $x\gg1$ in the second step. It should be noted that \eqref{fmapp} with $\gamma=0$ differs from \eqref{fmnon} by a $\sqrt{2\pi}\approx2.5$ pre-factor. This is because \eqref{fmapp} is obtained by the saddle point method which has assumed an approximate Gaussian peak. It may differ from the exact integrand by a factor, which is within $\mathcal{O}(1)$ when we make the comparison between \eqref{fmapp} and \eqref{fmnon}.

Therefore, we conclude that $\sigma^2 (R)$ can be uniquely reconstructed from $f(M)$
at least within the approximations made above.

\subsection{Step~(iii)}\label{s3}
Our final step is to reconstruct the mass function $f(M)$ from the merger rate density ${\cal R}(m_1, m_2, t)$ in the mass plane.
It is known that PBHs can form binaries either in radiation dominated era or at low redshift (e.g. \cite{Sasaki:2018dmp}).
In the former case, two neighboring PBHs which are accidentally closer than the mean distance between PBHs
become gravitationally bounded and the tidal force from the surrounding PBHs gives the initial 
angular momentum of the PBH binary.
In the latter case, two PBHs randomly moving inside the halo have a near miss at which they becomes
gravitationally bounded due to the energy loss by GW radiation.
If the fraction of PBHs in dark matter $f_{\rm PBH}$ is much smaller than unity,
which we assume in this paper, 
the former channel dominates the merger rate over the latter one.
The merger rate density in the former channel for the case of the extended PBH mass function has been studied in \cite{Raidal:2017mfl,Kocsis:2017yty, Raidal:2018bbj}.
In this paper, we adopt the most updated one \cite{Raidal:2018bbj} given by
\begin{align}\label{merger-rate}
\mathcal{R}(m_1, m_2, t)=\frac{1.6\times 10^6}{{\rm Gpc}^3 {\rm yr}}
f_{\rm PBH}^{\frac{53}{37}}
{\left( \frac{t}{t_0} \right)}^{-\frac{34}{37}}
{\left( \frac{m_t}{M_\odot} \right)}^{-\frac{32}{37}}
{\left( \frac{m_t^2}{m_1m_2} \right)}^{\frac{34}{37}}
S[f|f_{\rm PBH},m_t] m_1 m_2 f(m_1)f(m_2). 
\end{align}
Here $m_t=m_1+m_2$ and $S$, which is the functional of $f(m)$, 
is the so-called suppression factor which accounts for the distruption of the PBH binaries in later Universe. 
Now, our problem is to obtain $f(m)$ out of $\mathcal{R}(m_1, m_2, t)$ which is supposed to have been determined by observations.
If the dependence of the suppression factor on $m_t$ can be ignored, one can easily estimate the ratio of $f(m_1)$ and $f(m_2)$ as 
\begin{equation}\label{ratio}
\frac{f(m_1)}{f(m_2)}={\left( \frac{m_2}{m_1} \right)}^{\frac{21}{37}}
 \sqrt{\frac{\mathcal{R}(m_1,m_1,t)}{\mathcal{R}(m_2, m_2, t)}}.
\end{equation}
The normalization condition \eqref{normf1} gives the overall factor of $f(m)$.
Finally, $f_{\rm PBH}$ is obtained by substituting the obtained $f(m)$ into Eq.~(\ref{merger-rate}).
If the dependence of the suppression factor on $m_t$ is non-negligible, obtaining $f(m)$ requires more procedures.
From Eq.~(\ref{merger-rate}), we have
\begin{align}
\frac{f(m_1)f(m_2)}{f^2 \left( \frac{m_1+m_2}{2} \right)}
={\left( \frac{{\left( m_1+m_2 \right)}^2 }{4m_1 m_2} \right)}^{\frac{3}{37}}
\frac{\mathcal{R}(m_1,m_2,t)}{\mathcal{R}(\frac{m_1+m_2}{2},\frac{m_1+m_2}{2},t)}. \label{ratio-2}
\end{align}
In the following, we mainly concentrate on the case where the dependence of the suppression factor on $m_t$ is non-negligible.

We can construct $f(m)$ satisfying \eqref{ratio-2} in the following way. We first choose two masses $m_{1*}$ and $m_{2*}$.
($m_{1*} < m_{2*}$) arbitrarily and assign values given by hand to
$f(m_{1*})$ and $f(m_{2*})$.
Then, the above equation gives $f(\frac{m_{1*}+m_{2*}}{2})$.
From the values of $f(m_{1*})$ and $f(\frac{m_{1*}+m_{2*}}{2})$,
we can obtain $f(\frac{3m_{1*}+m_{2*}}{4})$. By repeating this procedure for $n$ times, we obtain the following recurrence relation
\begin{align}
&\frac{f(m_{1*})f(M_{n-1})}{f^2\left(M_n\right)}=\left(\frac{M_n^2}{m_{1*}M_{n-1}}\right)^{3/37}\frac{\mathcal{R}(m_{1*}, M_{n-1}, t)}{\mathcal{R}\left(M_n, M_n, t\right)}\,,\label{fmergern}\\
&M_n\equiv m_{1*}+\frac{\Delta m}{2^n}\,,\qquad \Delta m\equiv m_{2*}-m_{1*}\,,\qquad n=0,1,2,...\label{fmergern1}
\end{align}
where $M_0=m_{2*}$. From this relation, $f(M_n)$ can be further expressed as
\begin{align}\label{fmdis}
f(M_n)&=f(m_{1*})\left[\left(\frac{m_{2*}}{m_{1*}}\right)^{\frac{3}{37}}\frac{f(m_{2*})}{f(m_{1*})}\right]^{\frac{1}{2^n}}\left(\frac{m_{1*}}{M_n}\right)^{\frac{3}{37}}\prod_{i=1}^{n}\left[\frac{\mathcal{R}\left(M_{n-i+1}, M_{n-i+1}, t\right)}{\mathcal{R}(m_{1*}, M_{n-i}, t)}\right]^{\frac{1}{2^i}}\,.
\end{align}

\begin{figure}[t]
 \begin{center}
   \includegraphics[clip,width=10.0cm]{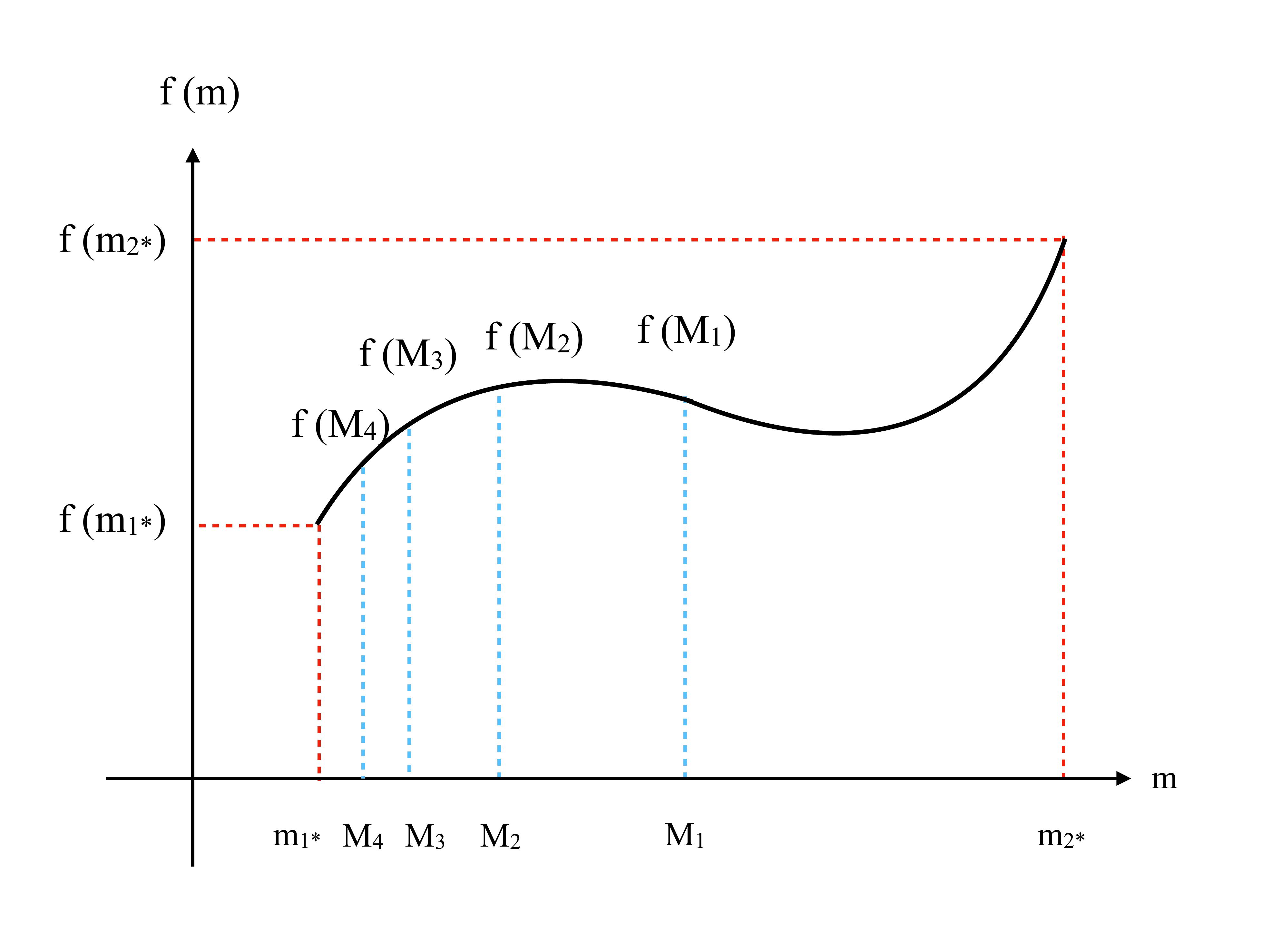}
   \caption{Illustration of the construction of the mass function $f(M_n)$. According to \eqref{fmergern1}, we divide the mass between $m_{1*}$ and $m_{2*}$ into $n$ pieces such as $M_1$, $M_2$, .... The corresponding mass functions, connected by the dashed blue lines, are then calculated by \eqref{fmdis}. For mass value $m\in(M_{i+1}, M_{i})$, we can set $f(m_1)=f(M_{i+1})$ and $f(m_2)=f(M_{i})$ in \eqref{ratio-2}, respectively, and repeat the above procedure to obtain the mass function for a more narrow mass range.}
   \label{fig-discrete}
 \end{center}
\end{figure}

As illustrated in Fig.~\ref{fig-discrete}, for mass value $m\in(M_{i+1}, M_{i})$, we can set $f(m_1)=f(M_{i+1})$ and $f(m_2)=f(M_{i})$ in \eqref{ratio-2}, respectively, and repeat the above procedure to obtain the mass function in a more narrow mass range. Therefore, provided with enough data of merger rate for different mass range, principally speaking, we can obtain the mass function in a sufficiently narrow mass range. 
In this case, the normalization condition \eqref{normf1} becomes the following form:
\begin{align}\label{normaldiscrete}
\sum_{n=0}^\infty f(M_n)\frac{\Delta m}{2^{n+1}}=1\,.
\end{align}

The value of $f(m)$ for $m$ outside $(m_{1*},m_{2*})$ can be obtained in a similar manner
by plugging $m_1 \to 2m_{1*}-m_{2*} <m_{1*}, \frac{m_1+m_2}{2} \to m_{1*}, m_2 \to m_{2*}$ 
(or $m_1 \to m_{1*}, \frac{m_1+m_2}{2} \to m_{2*}, m_2 \to 2m_{2*}-m_{1*} > m_{2*}$) in Eq.~(\ref{ratio-2}).
The solution $f(m)$ obtained in this way contains two free parameters (i.e. $f(m_{1*}), f(m_{2*})$)
which corresponds to the invariance of Eq.~(\ref{ratio-2})
under $f(m) \to Ae^{Bm} f(m)$ ($A,B$:constant).
One of the two parameters (we take it to be $A$) is fixed by the normalization condition (\ref{normaldiscrete}).
Thus, the general solution of Eq.~(\ref{ratio-2}) can be expressed as $f(m)=c(B)f_0(m)e^{Bm}$, where $f_0(m)$ is a particular solution
and $c(B)$ is the normalization constant.
Substituting this solution to Eq.~(\ref{merger-rate}) will enable us to fix $f_{\rm PBH}$ and $B$ simultaneously
since they enter differently in Eq.~(\ref{merger-rate}).

\section{A simple example: neglecting the effect of critical collapse and suppression factor}\label{example}
In this section, we consider the simplest case where the effects of the critical collapse and suppression factor are neglected. 
In such case, the primordial power spectrum can be analytically expressed in terms of the merger rate and other parameters, which we will show below.
We start with the assumption (0) that the function of merger rate density $\mathcal{R}(m, m, t)$ has been obtained from observation data. We then use step (iii) to determine the mass function. Let us choose a fixed mass point $m_*$ and design a value for $f(m_*)$. Since the effect of the suppression factor is neglected, using \eqref{ratio}, the mass function can be written as
\begin{align}\label{massfun1}
f(m)=g(t)m^{-\frac{21}{37}}\sqrt{\mathcal{R}(m, m, t)}\,,\qquad g(t)\equiv f(m_*)m_*^{\frac{21}{37}}\left[\mathcal{R}(m_*, m_*, t)\right]^{-\frac{1}{2}}\,.
\end{align}
It should be noted that $f(m)$ is independent from $t$ because of the normalization condition \eqref{normf1}. Secondly, from step (ii), since the effect of critical collapse is neglected here, the variance can be expressed in terms of the mass function by inverting \eqref{fmnon},
\begin{align}\label{sigmafun1}
\sigma(m(R))=\frac{\Delta_{\rm th}}{\sqrt{2}}\left[{\rm erfc}^{-1}\left(\frac{2f_{\rm PBH}\Omega_{\rm CDM}}{\left(M_{\rm eq}K^3\right)^{\frac12}}m^{\frac{3}{2}}f(m)\right)\right]^{-1}\,,
\end{align}
where ${\rm erfc}^{-1}(x)$ is the inverse function of the complementary error function defined by ${\rm erfc}\left[{\rm erfc}^{-1}(x)\right]=x$. We notice that in this simple case, the mass of PBH is approximated as the fraction of horizon mass such that $m=KM_R$. Hence, from \eqref{massR}, the mass of PBH is related to the horizon scale $k=1/R$ as~\cite{Green:2004wb}
\begin{align}\label{mass1}
m(k)=\left(\frac{k_{\rm eq}}{k}\right)^2M_{\rm eq}K\left(\frac{g_{*, \rm eq}}{g_*}\right)^{\frac13}\,,
\end{align}
where $g_*$ is the number of relativistic degrees of freedom which is expected to be of order $10^2$ in the early universe. At matter radiation equality, we have $g_{*, \rm eq}\approx3$ and $k_{\rm eq}= 0.01{\rm Mpc}^{-1}$. Inserting \eqref{massfun1} and \eqref{mass1} into \eqref{sigmafun1}, we obtain
\begin{align}\label{sigmafun2}
\sigma(m(k))=\frac{\Delta_{\rm th}}{\sqrt{2}}\left[{\rm erfc}^{-1}(A(m))\right]^{-1}\bigg|_{m=m(k)}\,,\qquad A(m)\equiv\frac{2f_{\rm PBH}\Omega_{\rm CDM}}{\left(M_{\rm eq}K^3\right)^{\frac12}}\left(g(t)m^{\frac{69}{74}}\sqrt{\mathcal{R}(m, m, t)}\right)\,.
\end{align}
Finally, we take step (i), i.e., insert \eqref{sigmafun2} into \eqref{relation-s-p}, then the primordial power spectrum can be obtained as
\begin{align}\label{powersim}
\mathcal{P}_{\cal R} (k)&=\frac{81}{16}\left(4\sigma^2+k\frac{{\rm d}\sigma^2}{{\rm d}k}\right)\nonumber\\
&=\frac{81}{8}\sigma^2\bigg[2-\sqrt{\pi}A(m)\left(\frac{69}{74}+\frac{m}{2\mathcal{R}(m, m, t)}\frac{{\partial}\mathcal{R}(m, m, t)}{{\partial}m}\right)\times\nonumber\\
&\qquad\times\left[{\rm erfc}^{-1}\left(A(m)\right)\right]^{\blue{-1}}\exp\left\{\left[{\rm erfc}^{-1}\left(A(m)\right)\right]^2\right\}\bigg]_{m=m(k)}\,,
\end{align}
where we have used the relation ${\rm d}\left[{\rm erfc}^{-1}(x)\right]/{\rm d}x=-(\sqrt{\pi}/2){\rm exp}\left(\left[{\rm erfc}^{-1}\left(x\right)\right]^{2}\right)$ in the last step. We notice that in such a simple case, once the parameters such as $f_{\rm PBH}$, $K$ and $\Delta_{\rm th}$ are known, the primordial powerspectrum is solely determined by the merger rate of the PBH binaries with the same mass.

\section{Conclusion and discussion}\label{conclusion}

In this paper, we have proposed the reconstruction method of primordial curvature perturbations on small scales through the merger rate of binary primordial black holes, which could form from large primordial curvature perturbation on small scales during radiation dominated era. This is motivated by the current observations of gravitational wave events by LIGO/VIRGO collaboration, which indicate high mass and low spin of the BBHs. These observations refresh the interest that PBHs may be one of the candidates of the corresponding BBHs.

Adopting the scenario that PBHs originate from the primordial curvature perturbations on small scales, we discussed how to reconstruct the power spectrum of primordial curvature perturbations from observations of (possible) PBH merger events, based on the following assumptions : 
  (0) In the future, some of the observed binary BH merger events
  are attributed to mergers of PBHs formed, and the merger rate distribution of the PBHs in the mass plane is observationally determined, (1) PBHs were formed out of rare high-$\sigma$ peaks of the primordial curvature perturbations in radiation dominated era, (2) The window function takes the {\it top-hat} form in $k$-space, (3) The primordial curvature perturbation follows {\it Gaussian} distribution and the effect of non-linearity between the curvature
  and the density perturbations is ignored. The reconstruction program is based on three steps. In the step (i), one could relate the primordial power spectrum $\mathcal{P}_{\cal R} (k)$ and the variance of the density perturbation $\sigma$ via \eqref{relation-s-p}. In the step (ii), we found the relation \eqref{fmapp} between the variance $\sigma$ and the PBH mass function $f(M)$ where the effect of the critical phenomena is included, given in \cite{Byrnes:2018clq,Wang:2019kaf}. In the step (iii), we presented the way to obtain the PBH mass function $f(M)$ from the merger rate ${\cal R}(m_1,m_2,t)$ of binary PBHs, adopting the theoretical expression of the merger rate given in \cite{Raidal:2018bbj}. By taking all steps, the power spectrum for the curvature perturbations can be uniquely determined by the PBH merger rate of binary black holes, provided with enough number of observed PBH merger events with various mass.

It should be noted that the violation of the assumptions (1)–(3) might not necessarily invalidate our methodology of the reconstruction.  For example, in the case of generic window function, the reconstruction of the primordial power spectrum from the variance is still possible as presented in the Appendix A.
Although the reconstruction procedures in this case becomes more complicated than in the case of the top-hat window function, the methodology for step (i) still holds. In this sense, the principle of our methodology is quite general though, once one relaxes each assumption, one needs to check whether there is one-to-one correspondence in each step of the reconstruction program.

On the other hand, there has been a long debate on which formalism should be used to calculate the PBH abundance and the mass function. Besides the formalism we have adopted in the paper, the peak theory formalism \cite{Bardeen:1985tr} is another one of the mostly used formalism which accounts the high-$\sigma$ peaks of the primordial perturbations for the PBH formation~\cite{Green:2004wb,Suyama:2019npc,Germani:2019zez,Yoo:2020dkz}. 
It remains to be addressed whether the reconstruction can be achieved in this case.

Finally, it is interesting to notice that the merger rate of BBHs with a same mass is needed in the reconstruction procedure, and one with different mass can be used for the consistency check. Therefore, this reconstruction of the power spectrum could be also a test of the PBH scenarios.
In fact, we can derive a consistency relation by using \eqref{merger-rate} as follows:
\begin{align}\label{Rm1m2}
\mathcal{R}(m_1, m_2, t)&=\left(\frac{m_1+m_2}{2\sqrt{m_1m_2}}\right)^{\frac{36}{37}}S[f|f_{\rm PBH},m_t]\sqrt{\frac{\mathcal{R}(m_1, m_1, t)\mathcal{R}(m_2, m_2, t)}{S[f|f_{\rm PBH},2m_1]S[f|f_{\rm PBH},2m_2]}}\,.
\end{align}
When the mass dependence of the suppression factor can be neglected, \eqref{Rm1m2} reduces to the consistency relation derived in Ref.~\cite{Kocsis:2017yty}. If the obtained mass function $f(m)$ determined only through $\mathcal{R}(m, m, t)$ cannot reproduce the merger rate for different masses adequately, that is, it does not satisfy the above consistency relation \eqref{Rm1m2}, it suggests us to include additional effects which are not included in \cite{Raidal:2018bbj}. Finally, even if we could include all of the relevant effects and the reconstruction program would not work, it might rule out the PBH scenarios. Thus, this reconstruction program is useful even to prove or to disprove the assumption (0).

\section*{Acknowledgements}

T. S. is supported by the MEXT Grant-in-Aid for Scientific Research on Innovative Areas No. 17H06359, and No. 19K03864. M. Y. is supported in part by JSPS Grant-in-Aid for Scientific Research Numbers 18K18764, and JSPS Bilateral Open Partnership Joint Research Projects. Y. Z. is supported by the Fundamental Research Funds for the Central Universities; the National Natural Science Foundation of China under Grants No. 12035011, No. 11975167, No. 11761161001, No. 11535004, and No. 11961141003; the National Key Research and Development Program of China under Grant No. 2018YFA0404403. This work was supported by Mitsubishi Foundation.

\appendix

\section{Reconstruction of ${\cal P}_{\cal R}(k)$ from $\sigma^2 (R)$}\label{appA}
The purpose of this appendix is to study how to reconstruct ${\cal P}_{\cal R}$
from $\sigma^2 (R)$ which is given by observations only for discrete values of $R$. 
Our starting point is Eq.~(\ref{sig2}) which gives $\sigma^2 (q)$ in terms of ${\cal P}_{\cal R}$
(Here, $q$ is wavenumber and we adopt the abuse notation $\sigma^2 (q) \equiv \sigma^2 (R=1/q)$ in stead of $\sigma^2 (R)$ for notational simplicity.).
We suppose observations have determined the values of $\sigma^2$ at discrete wavenumbers
${\bf Q}=(q_1,q_2,\cdots, q_N)$.
Since there are only $N$ measured values, we can determine the values of ${\cal P}_{\cal R}$ 
at maximally $N$ different wavenumbers.
At this point, there arises an ambiguity as to what combination of the $N$ different wavenumbers 
we choose for determining ${\cal P}_{\cal R}$ since, in general, the reconstructed ${\cal P}_{\cal R}$ will depend on the choice of the wavenumbers.
Here, we consider the natural choice that the wavenumbers for ${\cal P}_{\cal R}$ are also ${\bf Q}$
with the expectation that the ambiguities mentioned above become negligible for suficiently large $N$.
Under this prescription, the discretized version of Eq.~(\ref{sig2}) can be written as
\be
	\sigma^2 (q_i)=\frac{16}{81} \sum_{j=1}^N \frac{\Delta q_j}{q_j} W^2(q_j/q_i) {(q_j/q_i)}^4 {\cal P}_{\cal R}(q_j),
\ee
where $\Delta q_j$ is a suitable measure that approximates the integration whose explicit expression
depends on ${\bf Q}$.
These are the linear algebraic equations for $N$ variables ${\bf x}=({\cal P}_{\cal R}(q_1),\cdots, {\cal P}_{\cal R}(q_N))$.
Formally, they can be expressed by $N\times N$ matrix $M$ and the vector ${\bf S}$ as
\be
M {\bf x}={\bf S}, \label{linear-eq}
\ee
with the following identifications
\be
M_{ij}=\frac{16}{81} \frac{\Delta q_j}{q_j} W^2(q_j/q_i) {(q_j/q_i)}^4,~~~~~{\bf S}=(\sigma^2(q_1),\cdots,\sigma^2 (q_N)).
\ee
Thus, we can reconstruct ${\cal P}_{\cal R}$ within the discretization approximation if ${\rm dim}~{\rm Ker} M=0$.
In such as case, the reconstructed ${\cal P}_{\cal R}$ is given by
\be
{\bf x}=M^{-1} {\bf S}.
\ee
Whether $M^{-1}$ exists or not will in general depend on the shape of the window function as well
as the discretization of the wavenumbers, and in reality we have to check the existence of $M^{-1}$
in each case.

\end{document}